\documentclass[%
 reprint,
 superscriptaddress,
 amsmath,amssymb,
 aps,
prb,
]{revtex4-1}

\usepackage{graphicx}
\usepackage{dcolumn}
\usepackage{bm}
\usepackage{color}
\usepackage{xcolor}

\begin{document}


\title{Structuring co- and counter-flowing currents of polariton \\ condensates in concentric ring-shaped potentials}

\author{Franziska Barkhausen}
 \affiliation{Department of Physics and Center for Optoelectronics and Photonics Paderborn (CeOPP), Universit\"{a}t Paderborn, Warburger Strasse 100, 33098 Paderborn, Germany}
 
\author{Matthias Pukrop}%
\affiliation{Department of Physics and Center for Optoelectronics and Photonics Paderborn (CeOPP), Universit\"{a}t Paderborn, Warburger Strasse 100, 33098 Paderborn, Germany}%

\author{Stefan Schumacher}
\affiliation{Department of Physics and Center for Optoelectronics and Photonics Paderborn (CeOPP), Universit\"{a}t Paderborn, Warburger Strasse 100, 33098 Paderborn, Germany}%
\affiliation{Wyant College of Optical Sciences, University of Arizona, Tucson, AZ 85721, USA}%

\author{Xuekai Ma}
\affiliation{Department of Physics and Center for Optoelectronics and Photonics Paderborn (CeOPP), Universit\"{a}t Paderborn, Warburger Strasse 100, 33098 Paderborn, Germany}%

\date{\today}

\begin{abstract}
We investigate the current flow of microcavity polariton condensates loaded into concentric ring-shaped potentials. The tunneling of the condensates between different potential rings results in different phase-locked states, depending on the separation of the potential rings. As a consequence, the condensate currents in different rings can flow either in the same or opposite direction depending on the specific configuration of the ring-shaped potentials. In two concentric standard ring-shaped potentials, the condensates always circulate in the same direction (co-flowing current) and the vortices formed in the two rings share the same topological charge because of the azimuthally uniform distribution of their phase difference. In this case, increasing the number of the potential rings enables the excitation of Bessel-like solutions. If the two ring-shaped potentials are engineered into an eye shape with the inner ring being standard ring-shaped and the outer ring being elliptically ring-shaped, the phase differences of the condensates in the two rings along the major and minor axes of the ellipse can be opposite, which gives rise to a counter-flowing condensate currents. 
\end{abstract}

\maketitle


\section{\label{sec:Introduction}Introduction }
The exciton-polaritons studied in the present work are quasiparticles composed of photons and excitons in a planar quantum-well based semiconductor microcavity. As hybrid light-matter entities, exciton-polaritons can be excited optically and emit coherent light due to the finite polariton lifetime, which provides an excellent platform for structuring light on a few-micron length scale. Under excitation with a non-resonant optical beam, polaritons can still show macroscopic coherence~\cite{deng2002condensation,kasprzak2006bose}, also referred as non-equilibrium polariton condensates, enabled by efficient stimulated scattering of polaritons into the coherent phase. The repulsive polariton-polariton interaction also leads to the system showing strong nonlinearity, among other things driving spontaneous symmetry breaking and the formation of polariton vortices. Polariton vortices carry circular currents around a phase defect and can be created by various methods. For example, they can form spontaneously due to the inhomogeneity of the system and initial phase defects~\cite{lagoudakis2008quantized,lagoudakis2009observation,roumpos2011single} or the orbital angular momentum can be transferred directly from a coherent pump to the polaritons~\cite{sanvitto2010persistent,dominici2015vortex,PhysRevLett.116.116402}. Strong spatial anisotropy~\cite{PhysRevX.7.041006,PhysRevLett.121.085704} and optical lattices~\cite{PhysRevLett.101.187401} are also important factors to sustain polariton vortices. Recent research activities were much devoted to the structuring of the circular polariton currents by using diverse potential shapes, mainly focusing on ring-like potentials, including optically induced potentials~\cite{dreismann2014coupled,PhysRevLett.113.200404,liu2015new,PhysRevLett.120.065301,ma2020realization}, build-in potentials~\cite{PhysRevX.5.011034,PhysRevLett.121.225302,zezyulin2018spin,Barkhausen:20}, as well as the combination of them~\cite{PhysRevB.91.045305,PhysRevB.97.195149}.

Considering spatially more extended scenarios, the in-plane motion of polaritons and the polariton-polariton interaction give rise to the build-up of phase locked states for spatially separated condensates.~\cite{tosi2012geometrically,PhysRevLett.110.186403,PhysRevX.6.031032} These can either have symmetric or anti-symmetric phases, depending on their separation distance, geometrical arrangement\cite{PhysRevB.95.235301}, and outgoing polariton flows. A very large separation leads to the formation of the simple harmonic oscillator states~\cite{tosi2012sculpting}. This phase-locking influences not only the stationary fundamental mode, but also the higher-order modes that may feature flowing polariton currents. 

Polaritons loaded into multiple potential traps can also couple with each other due to quantum tunneling. In two overlapping micropillars, the tunneling of polaritons between the two pillars leads to the observation of Josephson oscillations~\cite{PhysRevLett.105.120403,abbarchi2013macroscopic}. In multiple overlapping pillars engineered to a hexagonal ring, the tunneling of photons between adjacent pillars and the polarization-dependent confinement allow the coupling between the spin and orbital momenta of polaritons~\cite{PhysRevX.5.011034}. Polariton condensates can also be excited to phase locked states in one-dimensional (1D) chain lattice potentials where the phase of the condensates can be either symmetric or antisymmetric depending on the pump intensity~\cite{PhysRevB.93.121303}. In such kind of 1D arrays, vortex chains can form, that is, the condensate in each pillar is excited to a vortex state, and their topological charges can be the same or opposite with similarity to the 1D spin systems with ferromagnetic and antiferromagnetic order~\cite{PhysRevLett.121.225302}. In our previous work, we found that in a single ring-shaped potential the phase coupling of condensates located at the centrosymmetric points supports multistable circular currents~\cite{Barkhausen:20}. If a potential barrier splits the ring-shaped potential to make it form a C-shape, polariton condensates in it exhibit pronounced coherent oscillations, which may persist far beyond
the coherence time of polariton condensates, passing periodically through clockwise and anticlockwise current states~\cite{xue2019split}. These may find applications in information processing, data storage, or implementation of quantum algorithms. 
 
In the present work, we use multiple concentric ring-shaped potentials to structure the current flow of polariton condensates in different potential rings on the basis of their phase coupling. Their phase difference can vary with the separation of the potential rings. We find that in two nested ring potentials the azimuthal isotropic phase difference of the condensates, which can be zero (symmetric) or $\pi$ (anti-symmetric), enables the generation of co-flowing currents. In the case of anti-symmetric coupling in multiple concentric ring-shaped potentials, Bessel-like solutions are found, that is, in each potential ring a vortex is excited and there is a $\pi$ phase jump between the neighboring vortices. Remarkably, if the phase difference of the condensates in different potential rings are azimuthally anisotropic, which can be realized in an eye-shaped potential with the inner part being a standard ring shape and the outer part being an elliptical ring shape, a counter-flowing condensate current in different rings can be excited. Due to the elliptical shape of the outer ring, the counter-flowing current is non-persistent, instead it oscillates robustly as time evolves and the oscillation period is much longer than the polariton lifetime.

\section{\label{sec:Model}Model}
\begin{figure} [t]
\centering
\includegraphics[width=1.0\columnwidth]{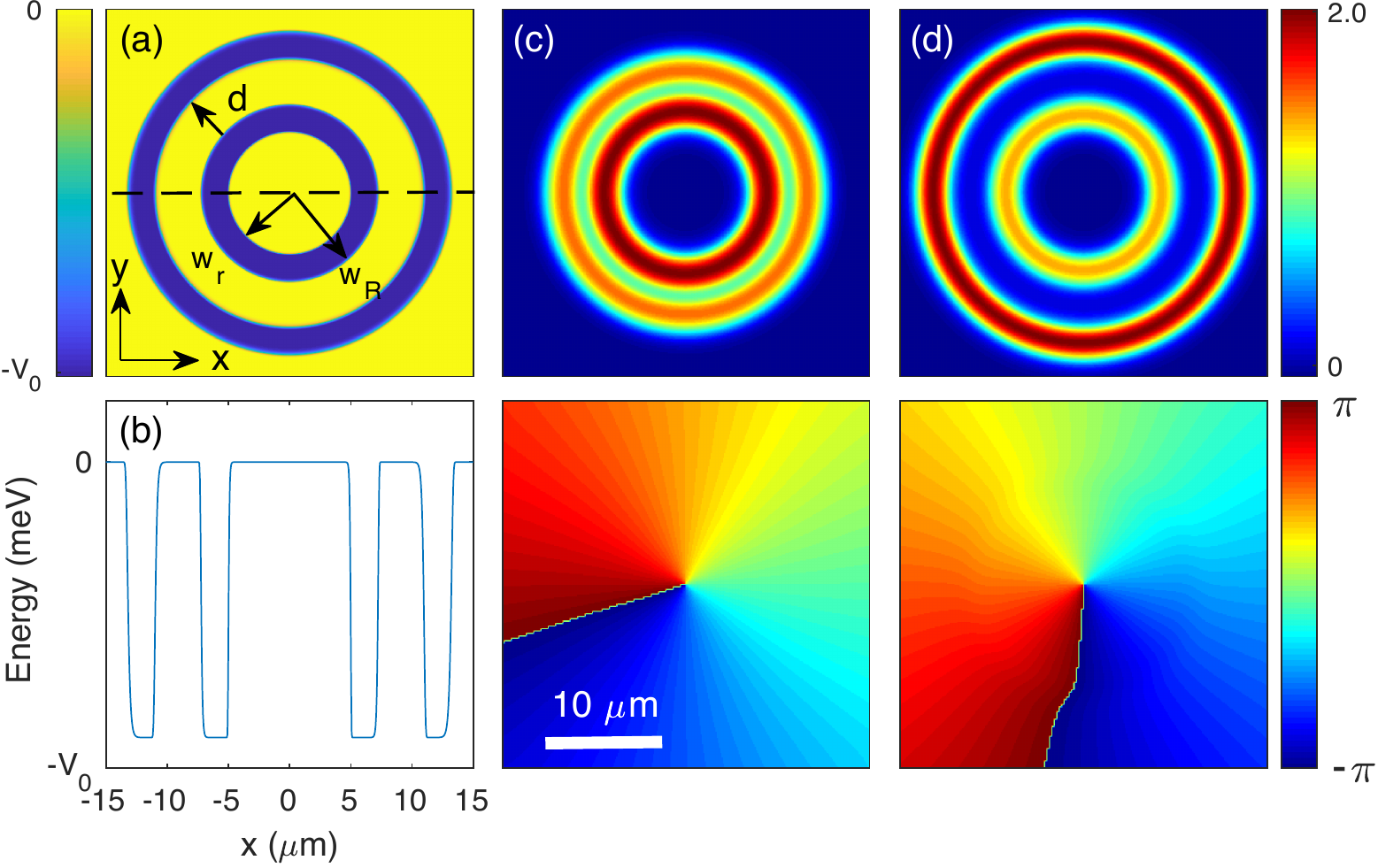}
\caption{{\bf Vortex steady states in two concentric ring-shaped potential wells.} (a) In-plane ($xy$-plane) landscape of the concentric ring-shaped potentials. $w_\text{r}$ and $w_\text{R}$ are the radii of the inner and outer edges, respectively, of the innermost well. The width of the ring is given by $w_\text{R}{-}w_\text{r}$. The separation of the rings (the distance from the outer edge of the inner ring to the inner edge of the outer ring) is denoted by $d$. (b) 1D distribution of the potential along the dashed line in (a). (c,d) Distributions of the density (top row) and the phase (bottom row) of the stationary vortex states with topological charge $m=1$ for (c) $d=1$ $\mu$m and (d) $d=3$ $\mu$m. Here, $w_\text{r}=5$ $\mu$m and $w_\text{R}=8$ $\mu$m.}\label{Fig1}
\end{figure}

To study the dynamics of polariton condensates in semiconductor microcavities under non-resonant excitation, we employ the driven-dissipative Gross-Pitaevskii equation coupled to an incoherent equation describing the density of the exciton reservoir~\cite{wouters2007excitations}:
\begin{equation}\label{e1}
\begin{aligned}
i\hbar\frac{\partial\Psi(\mathbf{r},t)}{\partial t}&=\left[-\frac{\hbar^2}{2m_{\text{eff}}}\nabla_\bot^2-i\hbar\frac{\gamma_\text{c}}{2}+g_\text{c}|\Psi(\mathbf{r},t)|^2 \right.\\
&+\left.\left(g_\text{r}+i\hbar\frac{R}{2}\right)n(\mathbf{r},t)+V(\mathbf{r},t)\right]\Psi(\mathbf{r},t),
\end{aligned}
\end{equation}
\begin{equation}\label{e2}
\frac{\partial n(\mathbf{r},t)}{\partial t}=\left[-\gamma_r-R|\Psi(\mathbf{r},t)|^2\right]n(\mathbf{r})+P(\mathbf{r},t)\,.
\end{equation}
Here, $\Psi(\mathbf{r},t)$ describes the coherent condensate field, and $n(\mathbf{r},t)$ describes the density of the reservoir. $m_{\text{eff}}{=}10^{-4}m_{\text{e}}$ ($m_{\text{e}}$ is the free electron mass) is the effective polariton mass around the bottom of the lower polariton branch which is assumed as parabolic. $\gamma_\text{c}{=}0.08~\mathrm{ps}^{-1}$ denotes the decay rate of the condensate and $\gamma_\text{r}{=}1.5 \gamma_c$ represents the decay rate of the reservoir. The polariton condensation rate is given by $R{=}0.01~\mathrm{ps}^{-1}~\mu\mathrm{m}^2$. The nonlinear coefficient $g_\text{c}{=}3\times 10^{-3}~\mathrm{meV}~\mu\mathrm{m}^2$ represents the strength of the polariton-polariton interaction and the strength of the polariton-reservoir interaction is given by $g_\text{r}{=}2 g_\text{c}$. The external potential is given by $V(\mathbf{r})$ with ring-shaped distribution and written as
\begin{equation}
V(\mathbf{r})=\sum_{j}-V_0e^{-\left(\textbf{r}'\right)^{2N}}(1-e^{-\left(\textbf{r}''\right)^{2N}})\,.
\end{equation}
Here, $\textbf{r}'=\mathbf{r}/(w_R+(j-1)(d+w_R-w_r))$, $\textbf{r}''=\mathbf{r}/(w_r+(j-1)(d+w_R-w_r))$, $V_0$ is the depth of the potential, $w_\text{r}$ and $w_\text{R}$ represent the radii of the inner and outer edges, respectively, of the innermost ring, and $d$ denotes the separation of the rings and it is defined as from the outer edge of the inner ring to the inner edge of the outer ring as indicated in Fig.~\ref{Fig1}(a). The index $j=$1, 2, 3, $\cdots$ represents the number of the potential rings. The integer index $N$ could be very large to make the potential very steep. Such kind of potential can be fabricated in planar semiconductor microcavities by different techniques~\cite{balili2007bose,lai2007coherent,wertz2010spontaneous,kim2013exciton,PhysRevB.93.121303}. The nonresonant optical pump $P(\mathbf{r},t)$ is a continuous-wave with a broad Gaussian shape and its spatial distribution satisfies
\begin{equation}
P(\mathbf{r})=P_{0}e^{-\mathbf{r}^2/w_g^2}.
\end{equation}
Here, $w_g=50$ $\mu$m, which is much larger than the size of the potential, and $P_0=1.1$ ps$^{-1}\mu$m$^{-2}$ (the threshold pump intensity $P_\text{thr}=$ 0.96 ps$^{-1}\mu$m$^{-2}$). In the following study, we keep the pump and the depth of the potential ($V_0$=0.45 meV) fixed and vary the radii of the ring-shaped potentials and their separations. In all the simulations in this work a vortex initial condition of the coherent condensate with topological charge $m=1$ and very small amplitude is applied. 

\section{\label{sec:tworing}Stationary and oscillatory dynamics of vortices in two concentric ring-shaped potentials}

We start from the simplest case with only two concentric ring-shaped potentials and each of them has a standard ring shape as shown in Fig.~\ref{Fig1}(a). In this case, the separation of the two rings is azimuthally isotropic, that is, $d$ is a constant. For a very small separation of the two rings with $d$=$1$ $\mu$m, the polariton condensates mainly occupy the inner ring as shown in Fig.~\ref{Fig1} (c). The phases of the condensates in both rings are fully synchronized due to the strong tunneling effect, thus they rotate in the same direction and act as a single condensate. When the separation is larger with $d$=$3$ $\mu$m, the second ring becomes mainly occupied [see Fig.~\ref{Fig1} (d)], although the pump intensity at the inner ring is stronger than that at the outer ring. This is because besides the circular motion, the condensate also propagates along the radial direction, supported by the gradient of the pump intensity, driving polariton density towards the region where the pump intensity is lower. Increasing the separation also weakens the quantum tunneling of the particles, resulting in slight desynchronization of the phases in the two rings as visible in Fig.~\ref{Fig1}(d).  

\begin{figure} [t]
\centering
\includegraphics[width=1.0\columnwidth]{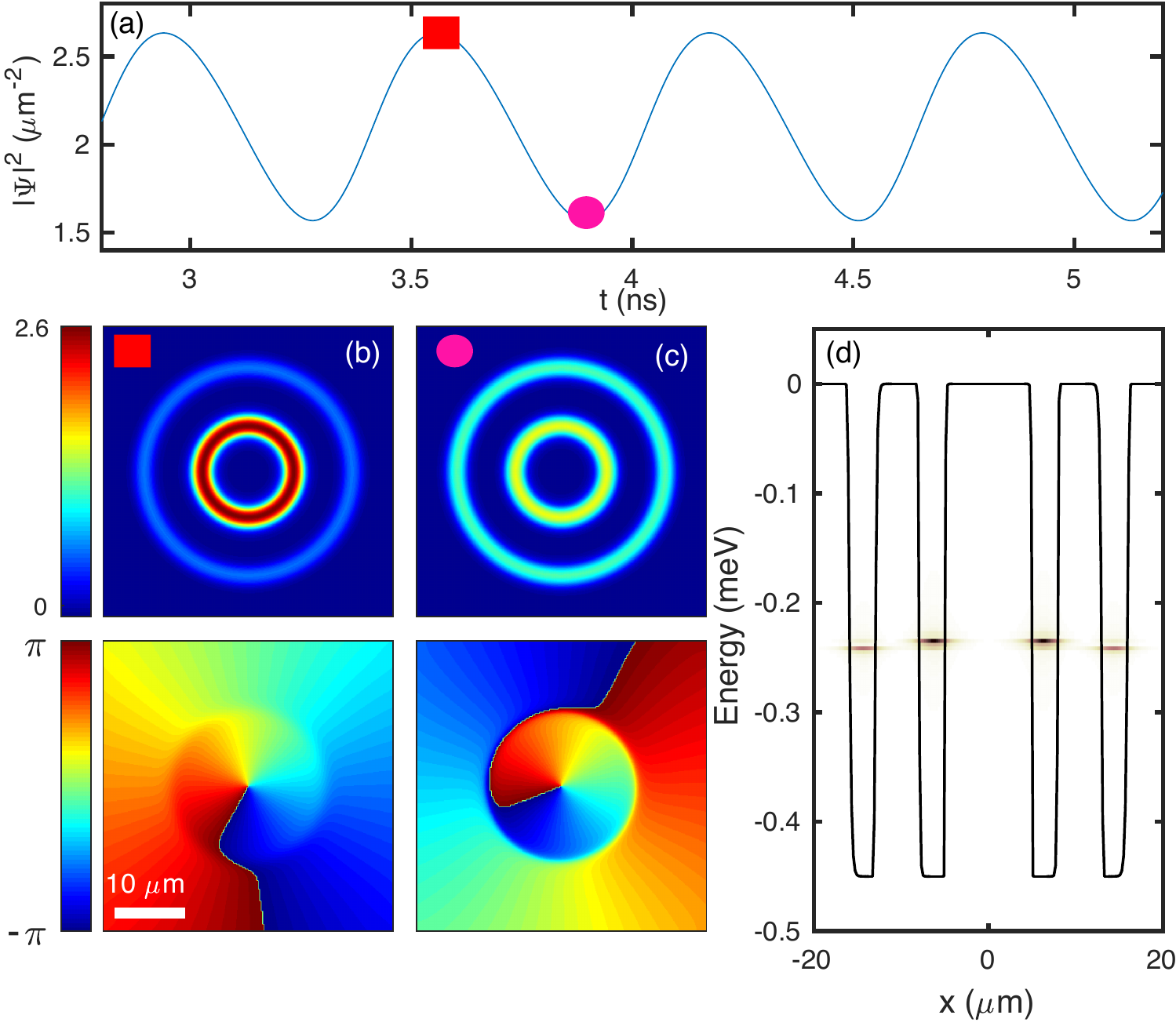}
\caption{{\bf Oscillatory dynamics of vortices in two concentric ring-shaped potentials.} (a) Time evolution of the peak density of the condensate. (b,c) Density (top row) and phase (bottom row) distributions of the oscillating vortex at different times, corresponding to the markers in (a), with a difference of $\Delta t=330$ ps. (d) Real-space spectrum of the oscillatory solution in (a) on top of the 1D distribution of the potential. Here, $w_\text{r}=5$ $\mu$m, $w_\text{R}=8$ $\mu$m, and $d=5$ $\mu$m.}\label{Fig2} 
\end{figure}

In principle, further increasing the separation of the two rings should lead to a $\pi$-phase jump between the condensates in different rings. However, the decrease of the pump intensity along the radial direction leads to a significant condensate density difference in the two rings as the separation becomes larger. Hence, the density-induced blue-shift results in a splitting into two distinct modes visible in the spectrum [Fig.~\ref{Fig2}(d)]. The beating of these two modes induces an oscillatory dynamics as shown in Figs.~\ref{Fig2}(a-c) for $d$=$5$ $\mu$m. At higher energy, the condensate in the inner ring rotates faster than the condensate in the outer ring as visible in the phase profiles in Figs.~\ref{Fig2}(b,c).

\section{\label{sec:Bessellike}Bessel-like vortices in four concentric ring-shaped potentials}
   
\begin{figure} [!b]
\centering
\includegraphics[width=1.0\columnwidth]{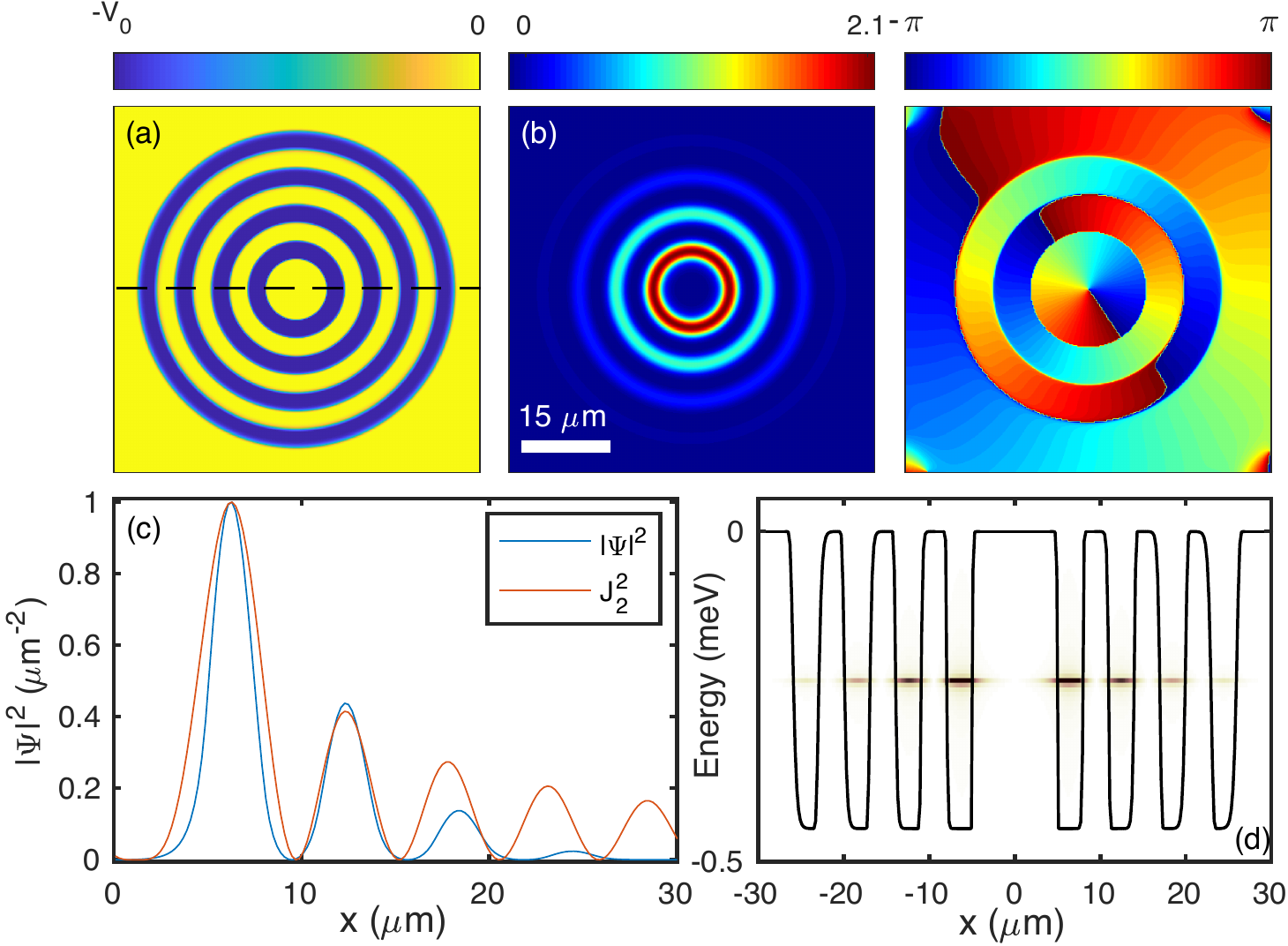}
\caption{{\bf Bessel-like vortices in four concentric ring-shaped potentials.} (a) In-plane landscape of the ring potentials. (b) Density and phase distributions of the Bessel-like vortex with $m=1$. (c) Comparison of the normalized Bessel mode $J_2$ with the normalized 1D solution selected from (b) along the dashed line at $y=0$ as indicated in (a). (d) Real-space spectrum of the solution in (b) on top of the 1D distribution of the potential along the dashed line in (a). Here, $w_\text{r}=5$ $\mu$m, $w_\text{R}=8$ $\mu$m, and $d=3$ $\mu$m.}\label{Fig3}
\end{figure}
 
\begin{figure*}
\includegraphics[width=2.0\columnwidth]{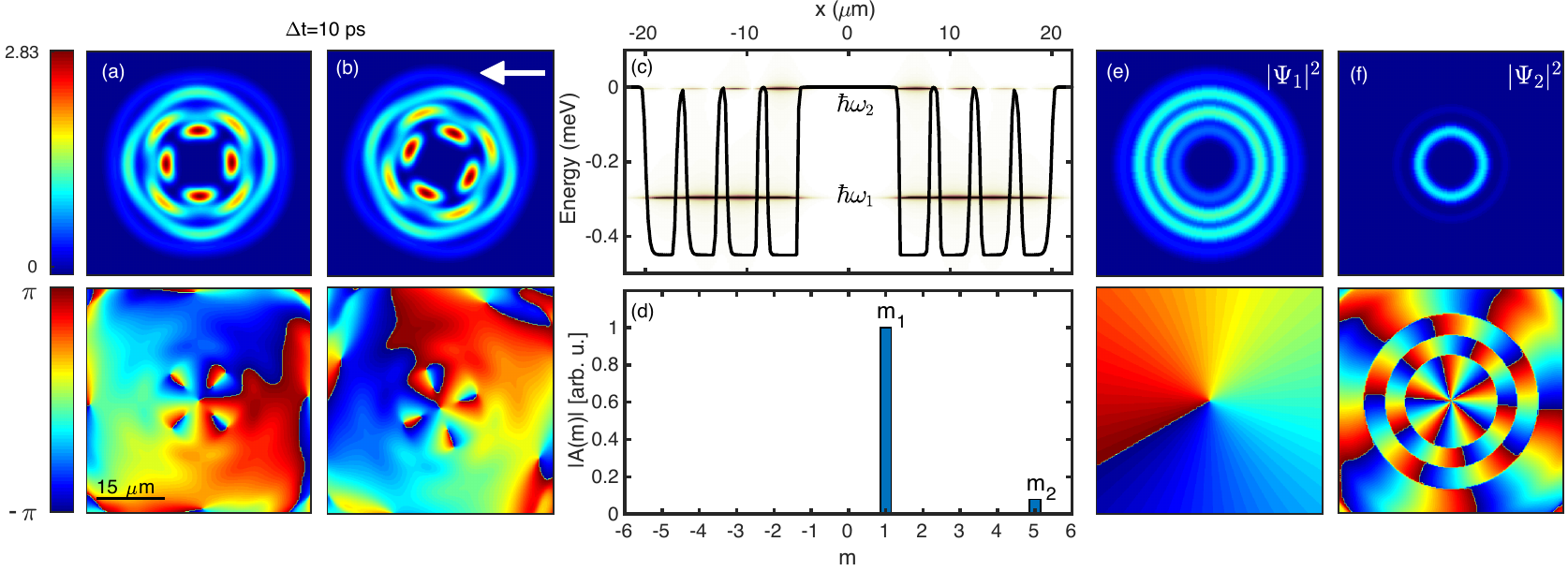}
\caption{\label{Fig4}{\bf Rotating solution in four concentric ring-shaped potentials.} (a,b) Density and phase distributions of the rotating solution at different times with $\Delta{t}=10$ ps (the period is 56 ps). (c) Real-space spectrum of the rotation solution on top of the 1D distribution of the potential. (d) Sorted distribution of the topological charges of the solution in (a). (e,f) Extracted density and phase profiles from the real-space spectrum in (c): (e) extracted from the fundamental mode $\Psi_{1}$ with $m_{1}=1$ and (f) extracted from the higher-order mode $\Psi_{2}$ with $m_{2}=5$. Here, $w_\text{r}=5$ $\mu$m, $w_\text{R}=8$ $\mu$m, and $d=1$ $\mu$m.}
\end{figure*}

As presented in Fig.~\ref{Fig1}(d), in a multi-ring configuration, most of the condensate is gathered in the outer ring due to the polariton outflow generated by the radially decreasing pump intensity. However, being trapped in a potential well, the polaritons cannot propagate further than the outermost ring. To study how the outgoing flow influences the distribution of the condensates in different rings, we extend the potential to four concentric rings as sketched in Fig.~\ref{Fig3}(a) and use the same parameters as in Fig.~\ref{Fig1}(d). From the solution in Fig.~\ref{Fig3}(b) one can see that the outflow of the condensates results in a small occupation of the outermost ring, although in that region the pump intensity is already below the condensation threshold. Remarkably, in this case, the phase between neighboring rings is not the same anymore, but instead a clear $\pi$-phase jump occurs. These $\pi$-phase jumps lead to a spatially separated arrangement of the vortices in different rings [see Fig.~\ref{Fig3}(d)]. The build-up of the $\pi$-phase jump is obviously caused by the appearance of more concentric rings in the potential by comparing with the solution in Fig.~\ref{Fig1}(d). The reason is that more rings enhance the tunneling effect and make the potential approaching a periodic one along the radial direction, recognized by the relatively smaller pump spot. It is known that in the 1D periodic potentials polariton condensates can be excited to the $\pi$-state (i.e., the boundary of the first Brillioun zone) when the pump intensity is slightly above the threshold~\cite{lai2007coherent,PhysRevLett.105.116402,tanese2013polariton,PhysRevB.93.121303}. Therefore, the same phase distribution shown in Fig.~\ref{Fig1}(d) vanishes for the configuration in Fig.~\ref{Fig3} under the same pump. The distribution of the solution in Fig.~\ref{Fig3}(b) is similar to that of a Bessel function of the first kind of second order $J_{2}$, as shown in Fig.~\ref{Fig3}(c). To more accurately create a Bessel vortex mode, the width and separations of the potential rings can be adjusted as needed (not shown).

\begin{figure*}
\includegraphics[width=2.0\columnwidth]{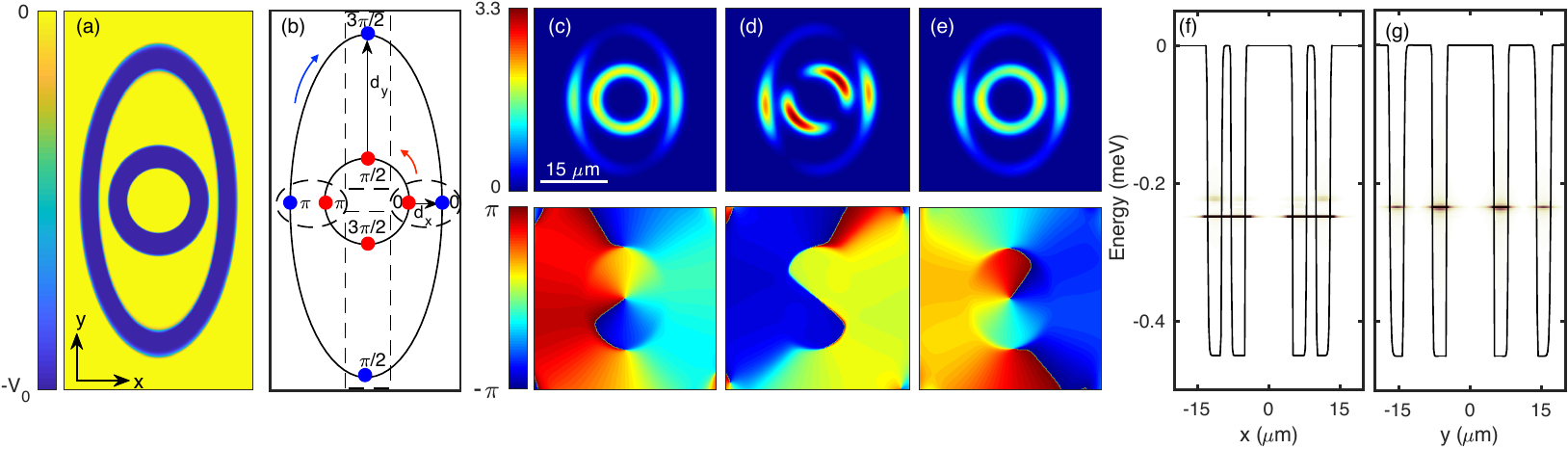}
\caption{\label{Fig5}{\bf Counter-flowing condensate currents in an eye-shaped potential.} (a) Landscape of the eye-shaped potential. (b) Schematic illustration of the counter-flowing condensate current. The separations of the two rings along the $x$ and $y$ directions are denoted by $d_x$ and $d_y$, respectively. Possible phase distributions of the condensates in the two rings are marked. The dashed ellipses enclose the points with the same phase, while the dashed rectangles enclose the points with $\pi$-phase difference. The red and blue arrows indicate the flow directions of the condensates in the inner and outer rings, respectively. (c-e) Density and phase profiles of the counter-rotating vortex at different times with $\Delta{t}=82.5$ ps (the oscillation period is 330 ps). Real-space spectra of the counter-rotating vortices selected along (f) the minor and (g) the major axes of the ellipse. The corresponding 1D potential distributions are superimposed. Here, $w_\text{r}=5$ $\mu$m, $w_\text{R}=8$ $\mu$m, $d_x=2$ $\mu$m, and $d_y=6$ $\mu$m.}
\end{figure*}

If the separation of the potential rings in Fig.~\ref{Fig3}(a) is reduced, the strong tunneling effect synchronizes their phases again and a rotating solution is formed as shown in Figs.~\ref{Fig4}(a,b) in which the density profile shows a wheel pattern and rotates with a period of 56 ps. During the rotation, the peak density of the solution remains constant. From the phase distribution and the real-space spectrum [Fig.~\ref{Fig4}(c)] one can see that the rotation results from the excitation of two different modes. The rotating states of polaritons in a ring-shaped potential can also be excited by a bichromatic pump~\cite{kartashov2019rotating}. The fundamental mode in Fig.~\ref{Fig4}(c) shows that the density along the radial direction is continuous, which means that the phases in the rings are the same. The higher-order mode, which carries a larger topological charge, at the edge of the potential well shows a spatially discrete spectrum along the radial direction, caused by the $\pi$-phase jumps that occur between neighboring rings as studied in Fig.~\ref{Fig3}(d). To find the phase information of each mode separately, in Fig.~\ref{Fig4}(a) we assume the solution $\Psi(\textbf{r},t)=\Psi_{1}(\textbf{r})e^{-i\omega_{1}t}+\Psi_{2}(\textbf{r})e^{-i\omega_{2}t}$ with $\Psi_{1}(\textbf{r})=\Psi_{01}(\textbf{r})e^{i\textbf{k}_{1}\textbf{r}}e^{im_{1}\phi}$ and $\Psi_{2}(\textbf{r})=\Psi_{02}(\textbf{r})e^{i\textbf{k}_{2}\textbf{r}}e^{im_{2}\phi}$, where $\Psi_{01}(\textbf{r})$ and $\Psi_{02}(\textbf{r})$ are the amplitude profiles, $m_{1}$ and $m_{2}$ are the topological charges, $\omega_{1}$ and $\omega_{2}$ are the frequencies, and $\phi$ is the polar angle. The contributions of the solution in Fig.~\ref{Fig4}(a) with topological charges $m$ can be extracted by projecting onto the respective orbital angular momentum components with:
\begin{equation}
A(m)=\int\Psi(\textbf{r})e^{-im\phi}d\textbf{r}.
\end{equation}
The result is shown in Fig.~\ref{Fig4}(d) where two topological charges are obtained: $m_{1}=1$ (corresponding to the fundamental mode) and $m_{2}=5$ (corresponding to a higher-order mode). To extract the density distribution $\Psi_{01}$ ($\Psi_{02}$) and the radial component of the phase $e^{i\textbf{k}_{1}\textbf{r}}$ ($e^{i\textbf{k}_{2}\textbf{r}}$) of the two modes, we Fourier transform separately each mode in Fig.~\ref{Fig4}(c) from the energy-spatial domain to the time-spatial domain, then select the resulting 1D profile at a fixed time scale and extend it to the 2D distribution by multiplying with the term $e^{im_{1}\phi}$ ($e^{im_{2}\phi}$). The two extracted profiles are shown in Figs.~\ref{Fig4}(e,f). The fundamental mode in Fig.~\ref{Fig4}(e) carries the topological charge $m_{1}=1$ and the phase is radially independent, similar to the result in Fig.~\ref{Fig1}(c). Due to the strong tunneling effect, the condensate propagates further away from the pump center and mainly resides in the two middle rings rather than in the innermost ring. For the higher-order mode in Fig.~\ref{Fig4}(f) at the edge of the potential, its innermost ring is dominated by the topological charge $m_{2}=5$ and $\pi$-phase difference between the neighboring rings. There are two factors that lead to the selection of $m_{2}=5$ as the contributing higher-order mode. The first one is that the initial condition is with topological charge $m=1$, which determines that the topological charge of the higher-order mode can only be an odd number (i.e., antisymmetric). The other one lies in the size of the innermost ring and the polariton lifetime. For a fixed polariton lifetime, the radius of the vortex ring and its topological charge are related~\cite{PhysRevB.93.035315}.

\section{\label{sec:counterrotation}counter-flowing condensate current in eye-shaped potentials}

Thus far, all the condensates in different rings propagate in the same direction. In other words, the vortices formed in different potential rings have the same topological charge. It is worth asking whether the condensates in different concentric ring-shaped potentials can propagate in opposite directions, that is, they carry opposite topological charges in neighboring rings. In this section, we analyze and demonstrate this kind of dynamics in an eye-shaped potential as sketched in Fig.~\ref{Fig5}(a) where the inner potential is of standard ring shape and the outer potential is of an elliptical ring shape. This configuration leads to an azimuthally anisotropic phase difference of the condensates between the two rings. For example, we assume that the separation of the two rings along the minor axis of the elliptical ring (the $x$ direction) $d_x$ is smaller than that along the major axis of the elliptical ring (the $y$ direction) $d_y$. In this case, the smaller $d_x$ induces that the condensates at the closest points [see the red and blue points within the dashed ellipses in Fig.~\ref{Fig5}(b)] in the two rings have the same phase. The larger $d_y$, however, leads to that the condensates at the farthest points [see the red and blue points within the dashed rectangles in Fig.~\ref{Fig5}(b)] in the neighboring rings can have a $\pi$ phase jump. If the condensate in the inner ring circulates counter-clockwise, as indicated by the red arrow, with topological charge $m=1$, the only possibility of the condensate in the outer ring is to circulate clockwise, as indicated by the blue arrow, and consequently its topological charge is $m=-1$ (here we neglect the higher-order vortices). Therefore, in this scenario, the condensates in the different rings of the eye-shaped potential can flow in opposite directions. 

The results of the polariton currents in the eye-shaped potential are shown in Figs.~\ref{Fig5}(c-d). The phase in Fig.~\ref{Fig5}(c) shows that the condensate in the inner ring propagates counter-clockwise, whereas the condensate in the outer ring propagates clockwise, evidencing the counter-flowing condensate current as discussed above. However, the currents of the condensates in the eye-shaped potential are not steady-states. After about 165 ps (half of the period) the currents stop simultaneously [Fig.~\ref{Fig5}(d)] and both of them start to rotate to the opposite directions [Fig.~\ref{Fig5}(e)]. The flipping of their topological charges repeats as time evolves, forming an oscillatory solution [a video showing the oscillatory dynamics in Fig.~\ref{Fig5} can be found in the Supplemental Material~\cite{videos}]. Importantly, the counter-rotation of the condensates is maintained during the oscillation. In the spectrum in  Fig.~\ref{Fig5}(f) one can see that along the minor axis of the elliptical ring the fundamental mode is excited with the same phase in the two rings, while along the major axis a $\pi$-phase state is excited as shown in Fig.~\ref{Fig5}(g). The energy difference along the perpendicular directions gives rise to the oscillatory dynamics. We note that this kind of counter-flowing condensate currents and their oscillations in an eye-shaped potentials are very robust against noise.

\begin{figure} [t]
\centering
\includegraphics[width=1.0\columnwidth]{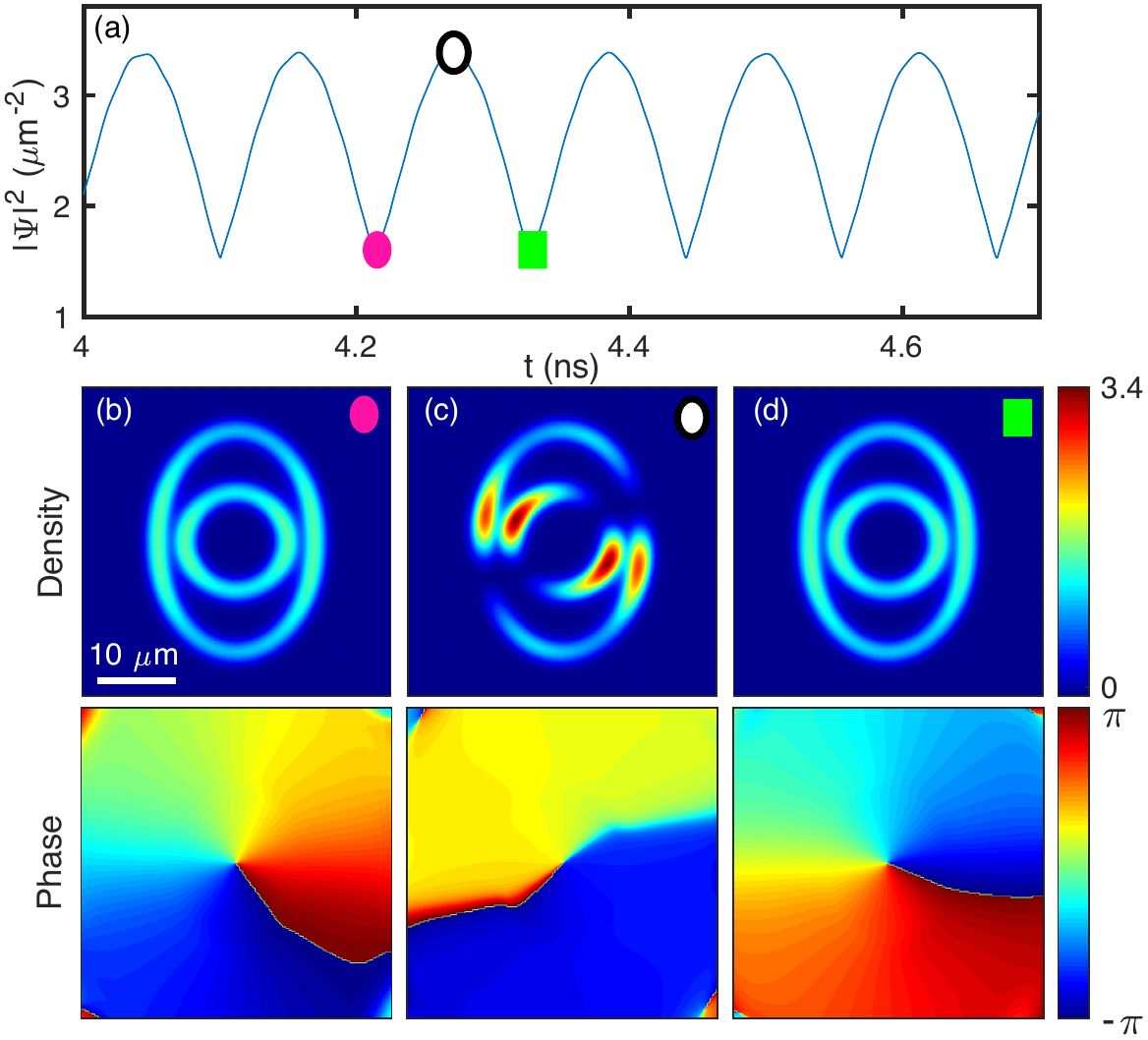}
\caption{{\bf Co-flowing condensate currents in an eye-shaped potential.} (a) Time evolution of the peak density of the condensate. (b-d) Density and phase profiles at different times corresponding to the marks in (a). Here, $w_\text{r}=5$ $\mu$m, $w_\text{R}=8$ $\mu$m, $d_x=1$ $\mu$m, and $d_y=5$ $\mu$m.}\label{Fig6}
\end{figure}

If we reduce the separation of the two rings in the eye-shaped potential along both the perpendicular directions, the tunneling effect is enhanced. The resulting stronger coupling of the condensates in the two potential rings leads to synchronization of their phases such that they possess the same topological charge as shown in Fig.~\ref{Fig6}. But, for the same reason as discussed above in Figs.~\ref{Fig5}(f,g), the energy difference along the $x$ and the $y$ directions still leads to an oscillation of the topological charge carried by the solution between $m=1$ and $m=-1$ as time evolves. We note that these oscillations in Figs.~\ref{Fig5} and \ref{Fig6} are fundamentally different from those we observed in our previous work where the oscillations of the topological charges were induced by the potential barriers~\cite{ma2020realization,xue2019split}.

{\section{\label{sec:conclusion}conclusion}}
We have successfully created co- and counter-flowing currents of polariton condensates loaded into a series of concentric ring-shaped potential wells. The nature of the current flow depends on the phase difference of the condensates in the different rings. If their phase difference is azimuthally invariant, for example in concentric standard ring-shaped potentials, the condensates propagate in the same direction and carry the same topological charge. If the phase differences along the perpendicular directions are opposite, for example in the eye-shaped potentials, the current of the condensate in different rings can propagate to the opposite directions. The counter-flowing currents in the eye-shaped potentials periodically and synchronously change their flow directions due to the energy difference along the major and minor axes of the elliptical potential. Our findings may be interesting for structuring light based on microcavity polaritons and may also trigger further investigations in non-concentric ring-shaped potentials.

\begin{acknowledgments}
This work was supported by the Deutsche Forschungsgemeinschaft (DFG) through the collaborative research center TRR142 (grant No. 231447078, project A04) and Heisenberg program (grant No. 270619725) and by the Paderborn Center for Parallel Computing, PC$^2$. X.M. further acknowledges support from the NSFC (No. 11804064).
\end{acknowledgments}



\nocite{*}

\providecommand{\noopsort}[1]{}\providecommand{\singleletter}[1]{#1}%

\end{document}